\documentclass[twocolumn,aps,prb,showpacs]{revtex4}
\usepackage{amsmath, amsfonts, bm}
\usepackage{graphicx}

\newcommand{\figwidth}{0.48\textwidth}

\begin{document}

\title{Quantum analogue of the spin-flop transition for a spin pair}

\author{B.~A.~Ivanov}
\email{bivanov@i.com.ua}

\author{V.~E.~Kireev}
\email{kireev@imag.kiev.ua}

\affiliation{Institute of Magnetism NAS of Ukraine, 36-B Vernadskii avenue,
03142 Kiev, Ukraine}

\date{\today}

\begin{abstract}
  Quantum (step-like) magnetization curves are studies for a spin pair with antiferromagnetic
  coupling in the presence of a magnetic field parallel to the easy axis of the magnetic anisotropy.
  The consideration is done both analytically and numerically for a wide range of the anisotropy
  constants and spins up to $S \gtrsim 100$.  Depending on the origin of the anisotropy (exchange or
  single-ion), the magnetization curve can demonstrate the jumps more than unity and the
  concentration of the unit jumps in a narrow range of the field.  We also point the region of the
  problem parameters, where the behavior is quasiclassical for $S = 5$, and where system is
  substantially quantum in the limit $S \to \infty$.
\end{abstract}


\maketitle

Mesoscopic magnetic systems whose total spin $S$ is large, but that exhibit quantum effects
associated with a finite value of the spin, are have been studied both experimentally an
theoretically in the last decade.  For instance, such systems are high-spin molecules with $S \sim
10$, magnetic clusters with $S \sim 100$, and magnetic dots (small magnetic particles with size 50
-- 100\,nm), see Ref.\cite{Wernsdorfer01, Skomski03}.  Objects that demonstrate quantum properties
in macroscopic (more accurate, mesoscopic) scales are important for the physics of the magnetism and
its applications: the implementation of the idea of a quantum computer\cite{Wernsdorfer+02nat,
LeuenLoss01nat} and information recording devices.\cite{Wernsdorfer01} The simplest quantum effect
that is manifested in such systems is associated with the quantization of the projection of the
total spin.  It consists in a step-like change of the magnetization when the external magnetic field
changes continuously.\cite{Wernsdorfer01, Skomski03} Such a behavior is observed for many systems
with an antiferromagnetic (AFM) interaction, see for a review.\cite{ShapiraBind02} The most studied
systems are spin dimers with AFM interaction, for instance, pairs of high-spin molecules Mn$_4$ with
the maximum total spin $2S = 9$, see\cite{Wernsdorfer+02prb, Edwards+03} Measurements are also
carried out for spin triplets, quartets, etc.\cite{ShapiraBind02, Waldmann00} The analysis of jumps
on the magnetization curves is useful for experimental determination of material
constants.\cite{ShapiraBind02}

For realization of this method, it is important to note that if anisotropy terms are absent in the
Hamiltonian~\eqref{ham}, i.e. $B = 0$ and $\kappa = 0$, the problem has an exact solution for an
arbitrary value of the spin.  The eigenstates of the Hamiltonian are states with a fixed total spin
$S$ and its projection $S^z$; for a constant $S$ the state with the minimum energy correspond to the
maximum of the projection $S^z$, $S^z = S$.  The energy can be written as $E(S, S^z) = J S (S + 1) /
2 - g \mu_B H S^z$, and $S^z$ changes discontinuously with jumps from $S^z = n - 1$ to $S^z = n$ at
the field $H_n = J n / (g \mu_B)$; The saturated state is reached at $H = H_{ex}$, where the
exchange field is $H_{ex} = 2J S / (g \mu_B)$.  Other exactly solvable models for clusters with AFM
interaction are discussed in the review.\cite{Waldmann00} In the presence of magnetic anisotropy,
the quantum problem has no exact solution, excluding Ising model that corresponds to $\kappa = 1$
and $B = 0$ in the Hamiltonian~\eqref{ham}.  In the latter case, which is classical in essence, $S^z
= 0$ at $H < H_{ex}$, and the saturated state $S^z = 2S$ at $H > H_{ex} = J S / (g \mu_B)$.

We study the magnetization curves for a spin pair with antiferromagnetic coupling and uniaxial
anisotropy in the magnetic field directed along the easy axis.  It is found that the magnetization
curve appeared to be more complicated than for the case of a pure exchange interaction: the jumps
become nonequidistant and their value $\Delta S^z$ can exceed unity.  The behavior of the
magnetization curve also depends on the origin of the anisotropy.  It is noted similarity of a spin
pair and a macroscopic AFM close to the point of the spin-flop transition (SFT), that it can be
useful for the qualitative study of the system.  We also find out values of the field where
discrepancies between the quantum and classical model are not small even in the limit $S \to
\infty$.

\section{Quantum model and its quasiclassical analysis} 

Let us consider the Hamiltonian for a pair of spins $\bm{S}_1$ and $\bm{S}_2$ with antiferromagnetic
interaction $J > 0$, uniaxial anisotropy, and an external magnetic field $\bm{H}$ directed along the
easy axis $z$:
\begin{multline}\label{ham}
\mathcal{H} =
J [S_1^z S_2^z + (1 - \kappa)\bm{S}_1^\perp \bm{S}_2^\perp] - \\
-\frac{B}{2} [(S_1^z)^2 + (S_2^z)^2]
-g \mu_B H (S_1^z + S_2^z) \;.
\end{multline}
Here, both the single-ion anisotropy $B$ and the exchange anisotropy $\kappa J$ are taken into
account.  We start from the classical analysis of the problem.  If the operators $\bm{S}_1$ and
$\bm{S}_2$ are considered as classical vectors (spin of sublattices) of a macroscopic sample, a
useful resemblance with the problem of SFT appears, see Ref.\cite{BorRom62, Belov+79, BarIv85}.
Denote $\bm{l} = (\bm{S}_1 - \bm{S}_2) / |\bm{S}_1 - \bm{S}_2|$; in the polar parametrization $l_z =
\cos\theta$, $l_x + il_y = \sin\theta \exp(i\phi)$.  Eliminating the slave variable $\bm{S}_{tot} =
\bm{S}_1 + \bm{S}_2$ (the vector of the total spin)
\begin{equation}\label{Sz-slave}
\bm{S}_{tot} = 
\frac{2 g \mu_B H(\hat{\bm{e}}^z - \bm{l} \cdot \cos\theta )}
{J (2 - \kappa + \beta\cos 2\theta)} \;,
\end{equation}
where $\beta = B / J$ and $\hat{\bm{e}}^z$ is the unit vector along the symmetry axis, see details
in Ref.\cite{BarIv85}, the energy~$\mathcal{W}(\theta)$ can be written as
\begin{equation}\label{energy}
\mathcal{W}(\theta) =
-\frac{(g \mu_B H)^2 \sin^2\theta}{J (2 - \kappa + \beta \cos 2\theta)} +
J(\kappa + \beta)S^2 \sin^2\theta \;.
\end{equation}

In the absence of the magnetic field a standard result is renewed: both exchange and single-ion
anisotropy make additive contributions in the effective magnetic anisotropy $(\kappa + \beta) J S^2
\sin^2\theta$ that is easy-axis one at $\kappa + \beta > 0$.  Minimizing this energy at $H \neq 0$,
we obtain various magnetic phases.  The collinear phase $\Phi_\parallel$ is stable at $H < H_1$ and
$S^z = 0$ ($\theta = 0$ or $\pi$) in it.  The spin-flop phase $\Phi_\perp$ is stable at $H > H_2$.
It is characterized by $\theta = \pm\pi / 2$, and the projection of the total spin linearly depends
on the field, $S^z = 2 S H / H_{ex}$.  When $H \geq H_{ex} = J S (2 - \kappa - \beta) / (g \mu_B)$
the saturated state with $S^z = 2S$ exists.  Characteristic fields separated the regions of
stability are determined as
\begin{align}\label{H1H2}
H_1 & =
\frac{J S}{g \mu_B} \sqrt{(2 - \kappa + \beta) (\kappa + \beta)} \;, \quad
H_2 =
\frac{H_1^2}{H_{sf}} \;, \\
H_{sf} & =
\frac{J S}{g \mu_B} \sqrt{(2 - \kappa - \beta) (\kappa + \beta)} \;.
\end{align}
It is note that in the absence of the single-ion anisotropy the expression for the exchange field
coincides with the saturation field for the exactly solvable quantum models of the Heisenberg and
Ising magnets.

If the anisotropy is purely exchange, $\beta = 0$, the degenerate case with the equal fields $H_1$
and $H_2$ appears.  If $\beta > 0$, the transition from the collinear state $\Phi_\parallel$ to the
spin-flop state $\Phi_\perp$ is discontinuous (SFT is a first-order transition).  The respective
field $H = H_{sf}$ is introduced in the Eqs.~\eqref{H1H2}.  The value $S^z$ is changed on $\Delta
S^z_{cl}$, where
\begin{equation}\label{dsz}
\Delta S^z_{cl} =
2 S \sqrt{\frac{\kappa + \beta}{2 - \kappa - \beta}} \;.
\end{equation}

\begin{figure}
\includegraphics[bb = 65 230 535 790, width = \figwidth]{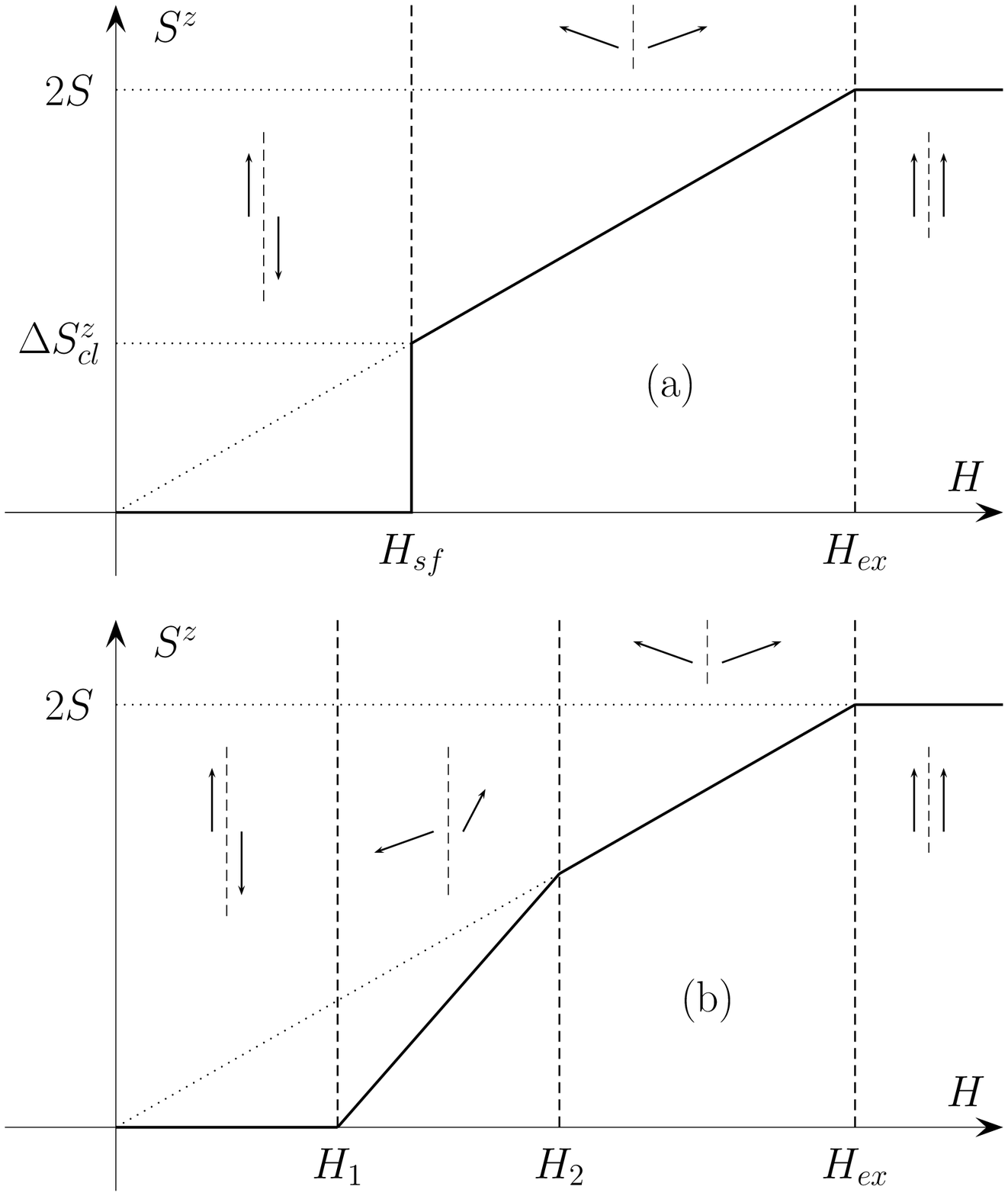}
\caption{The ground state of the classical AFM with single-ion and exchange anisotropies for
  different values of the field (schematically).  The arrows are spin directions in the different
  states of AFM. (a) --- the single-ion anisotropy prevails; (b) --- the exchange anisotropy
  prevails.
 \label{f:phases}}
\end{figure}

It is possible another case $H_1 < H_2$, when the transition $\Phi_\parallel \leftrightarrow
\Phi_\perp$ passes via two second-order transitions with a skew phase $\Phi_\angle$ between
them.\cite{Belov+79, BarIv85} The model~\eqref{energy} require the competitive anisotropies --- the
single-ion one with $\beta < 0$, that at $\kappa = 0$ lead to easy-plane anisotropy, and a large
exchange anisotropy $\kappa > -\beta$.  The effective anisotropy can be easy-axis.  In the range of
fields $H_1 < H < H_2$ the value $S^z$ changes linearly from 0 to $\Delta S^z |_{H = H_2} \approx
\Delta S^z_{cl}$, but the slope of the straight line is deeper than in the case of the spin-flop
phase, see Fig.~\ref{f:phases}.

\section{Quantum properties}

These results can be juxtaposed with the magnetization curves obtained from the quantum
model~\eqref{ham} in the limit case of the large spin if we adopt as a hypothesis that all linear
dependencies $S^z(H)$, see Fig.~\ref{f:phases}, are changed to the step-like ones with equidistant
unit jumps $\Delta S^z = 1$, and the jump $\Delta S^z_{cl} > 1$ at the point of SFT holds in the
quantum case.  This is reasonable when the field of SFT is stronger than the field of the first
quantum jump $H_{sf} > J / (g \mu_B)$ for the isotropic model, and $\Delta S^z_{cl} > 1$.  Both
conditions lead to the inequality $\kappa + \beta > 1 / (2 S^2)$.  It is valid in the limit case $S
\gg 1$ even for small anisotropy.

\begin{figure}
\includegraphics[bb = 65 500 535 790, width = \figwidth]{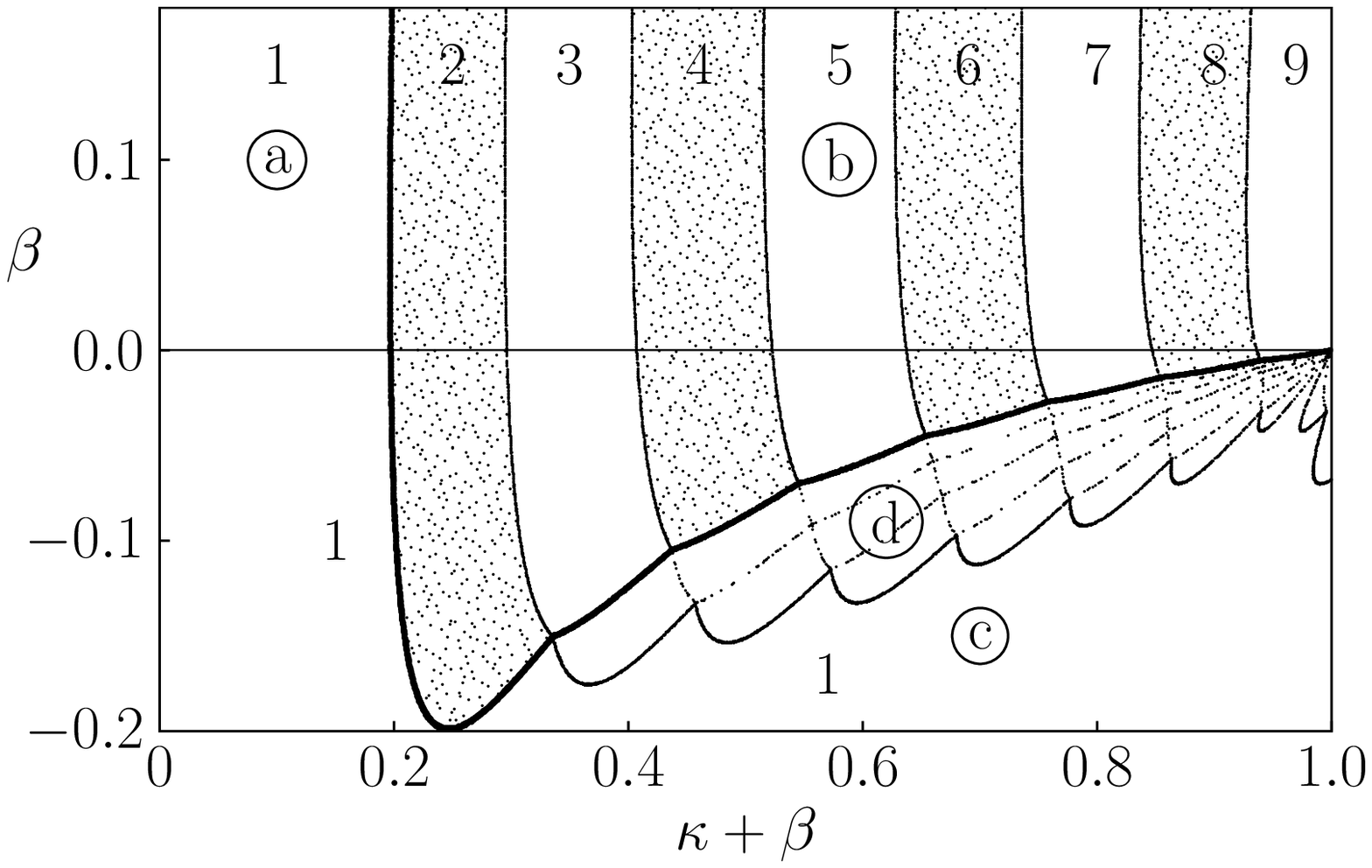}
\caption{Regions with a different behavior of the magnetization curve in the plane $(\kappa + \beta,
  \beta)$ for $2S = 10$.  The lines separate the regions of the plane with different values of the
  first jump $\Delta S^z$ (indicated by digits).  Regions with even $\Delta S^z$ are shaded.  The
  jump with $\Delta S^z = 10$ occurs when $\kappa + \beta \geq 1$; the only point (the Ising point
  $\kappa = 1$, $\beta = 0$) correspond it in the figure.  Letters in circles mark the
  characteristic regions of the phase diagram; the behavior of the magnetization $S^z(H)$ for them
  is presented in Fig~\ref{f:jumps}.  Bold curve separates region where $\Delta S^z > 1$.  The curve
  below it limits the region with lattice of parallelograms where the first jump is $\Delta S^z =
  1$, but among subsequent jumps the value $\Delta S^z > 1$ is present, see Fig.~\ref{f:jumps}(d)
  and explanation in the text.
 \label{f:first-jumps}}
\end{figure}

It is convenient to depict the behavior of the system in the $(b, \beta)$ plane, where $b = \kappa +
\beta$ is the constant of the effective anisotropy, see Eq.~\eqref{energy}.  According to the
classical consideration, SFT with the jump $\Delta S^z_{cl} > 1$ can occurs only for $\beta > 0$,
and regions with $\Delta S^z_{cl} = n > 1$ are vertical stripes bounded from the bottom by the
straight line $\beta = 0$ and from the sides by the vertical lines $b = b_n$ and $b = b_{n + 1}$,
where $b_n = 8 n^2 / (n^2 + 4 S^2)$.  The behavior with equidistant unit jumps similar to the
isotropic model can also be expected for the cases of a small anisotropy or an easy-plane anisotropy
$b < 0$.  For large values of anisotropy $b$ and $\beta < 0$ the value $S^z = 0$ persists up to the
field $H_1$, after that the concentration of $n$ jumps takes place in the narrow range $H_1 < H <
H_2$.  Thus, the classical approach gives a rich variety of types of the magnetization curves.  This
prediction is corroborated by the numerical results obtained for the quantum model~\eqref{ham}, but
an important exception must be made.

\begin{figure}
\includegraphics[bb = 65 350 265 790, width = \figwidth]{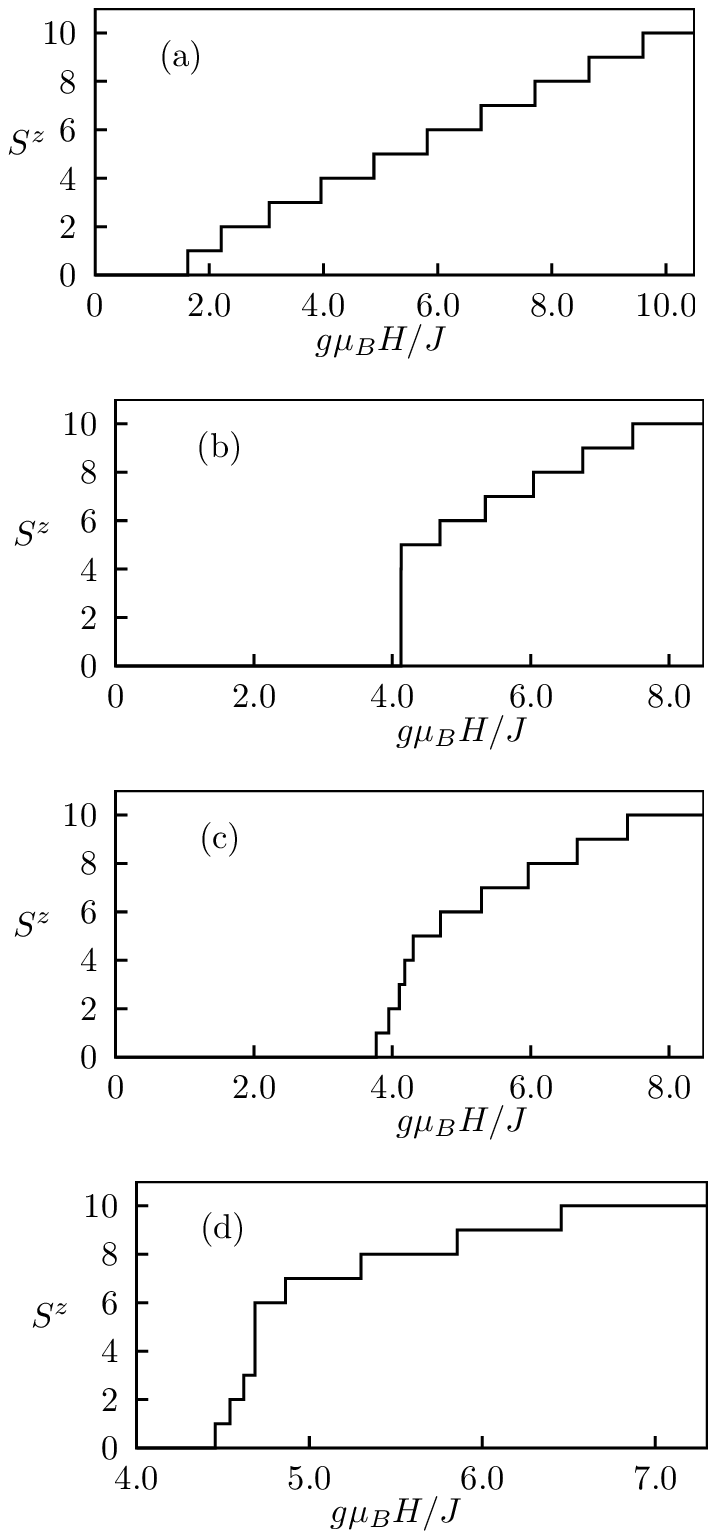}
\caption{Dependence $S^z$ on the field for the four characteristic regions denoted by letters in
  Fig.~\ref{f:first-jumps}.  Data is obtained for $2S = 10$. (a) --- almost exchange behavior: $b =
  0.1$, $\beta = -0.2$; (b) --- merging of jumps, an analogue of SFT: $b = 0.515$, $\beta = 0.1$;
  (c) --- strong concentration of jumps, an analogue of a skew phase $\Phi_\angle$: $b = 0.56$,
  $\beta = -0.1$; (d) --- nonunit jump $\Delta S^z > 1$ from state with $S^z \neq 0$, an analogue of
  the phase transition $\Phi_\angle \leftrightarrow \Phi_\perp$: $\beta = -0.09$, $b = 0.7$, see
  detailed description in Fig.~\ref{f:pt}.
 \label{f:jumps}}
\end{figure}

\section{Numerical simulation}

The Hamiltonian~\eqref{ham} commutes with the operator of the total spin projection $\hat{S}^z =
\hat{S}_1^z + \hat{S}_2^z$, and the complete Hilbert space of the problem decomposes into subspaces
with fixed values of $S^z$.  The operator~\eqref{ham} projected onto these subspaces has the form of
three-diagonal matrices whose eigenvalues can be numerically found using the QR-algorithm.  Another
important simplification of the problem is possible because the Zeeman term commutes with
$\hat{S}^z$, and the eigenvalues of the Hamiltonian~$E(S^z, 0)$ obtained for the case $H = 0$ can be
used for any field through the shift, $E(S^z, H) = E(S^z, 0) - g \mu_B H S^z$.  These properties
permit analyze the problem in the complete phase plane $(\kappa, \beta)$, even for a sufficiently
large values $2S \sim 250$ in a reasonable time on a personal computer.

Numerical investigation shows that almost all properties of the classical SFT hold in the quantum
model~\eqref{ham}, even for a small value of spin $S \sim 5$.  First, the state with $S^z = 0$
exists in the range $0 < H < H_1$.  Second, jumps with $\Delta S^z > 1$ are also observed for $\beta
> 0$, when the classical SFT is a first-order transition.  Third, the concentration of jumps in the
narrow range $H_1 < H < H_2$ is observed.  The shape of regions with a fixed value $\Delta S^z > 1$
qualitatively corresponds to the classical calculations even for small spins.  Particularly, the
curves separated regions with different $\Delta S^z = n > 1$ at $\beta > 0$ are approximatively
vertical, see Fig.~\ref{f:first-jumps}.  But qualitative discrepancies are also evident.  First of
all, for small $\Delta S^z$ regions with $\Delta S^z > 1$ extend into the lower half-plane.

\begin{figure}
\includegraphics[bb = 65 460 535 790, width = \figwidth]{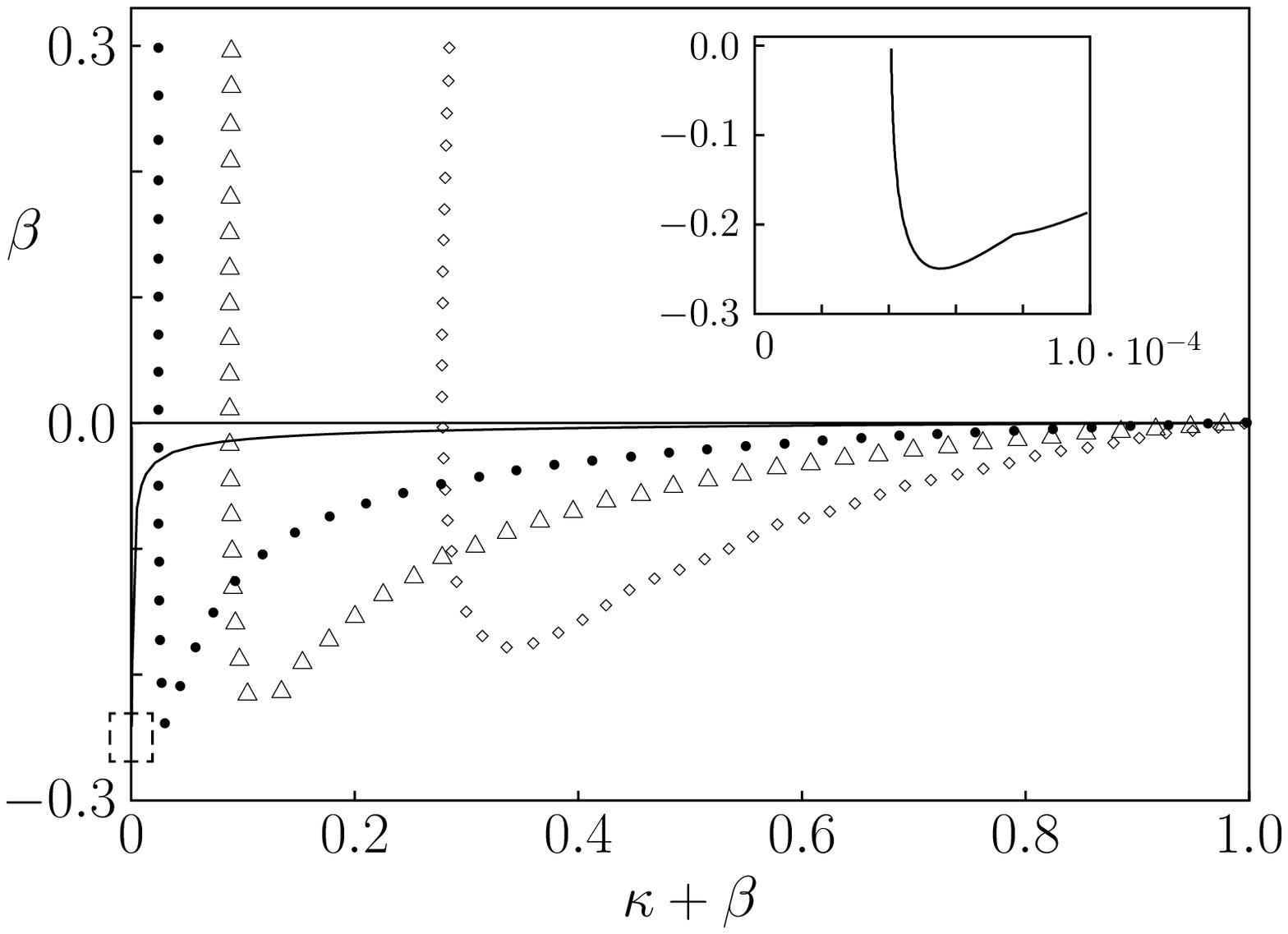}
\caption{Curves separating the spin-flop and skew phases for three values of the spin: $\diamond$ --
  $2S = 8$; $\triangle$ -- $2S = 16$; $\bullet$ -- $2S = 32$; the continuous line denotes data for
  $2S = 256$.  The dashed rectangle is the region where the minimum of the continuous line occurs.
  In the inset --- this region and the part of the curve adjacent to the $y$-axis for $2S =
  256$. \label{f:seps}}
\end{figure}

The behavior of the line separating regions with $\Delta S^z = 1$ and $\Delta S^z > 1$ is
astonished.  The discrepancy from the classical result $\beta = 0$ is not small even for large spins
$2S \simeq 200$, see Fig.~\ref{f:seps}.  The extent of this region in the direction of $\beta$ grows
when the spin increases and arrives at the limit value $\beta \sim -0.3$, but its area decreases
with $S \to \infty$.

\begin{figure}
\includegraphics[bb = 65 460 550 790, width = \figwidth]{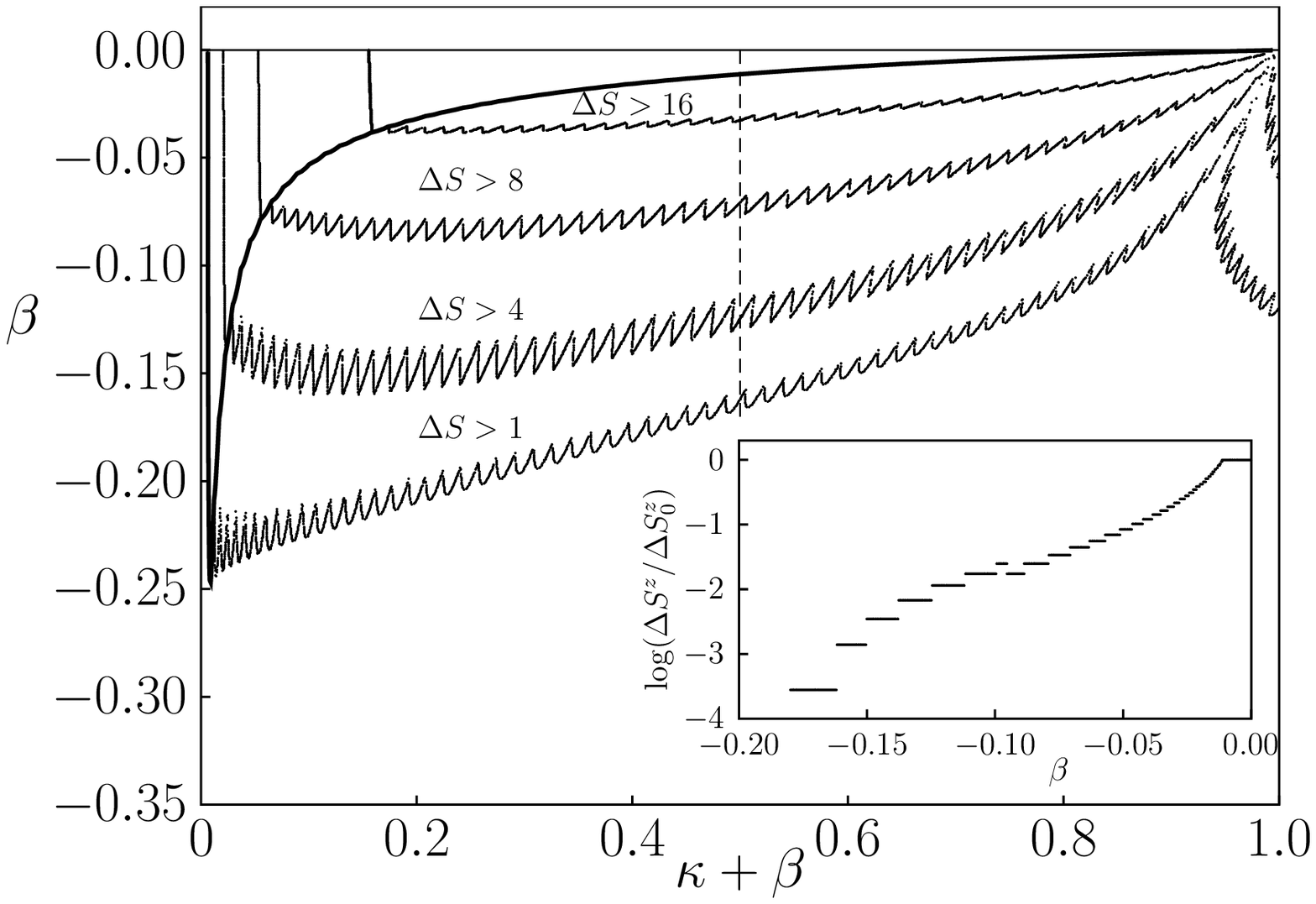}
\caption{Curves separating regions with a fixed jump value in the transition $\Phi_\angle
  \leftrightarrow \Phi_\perp$ for $2S = 64$.  Above the bold line the jump $\Delta S^z > 1$ occurs
  from a state with $S^z = 0$, see correspondent lines in Figs.~\ref{f:first-jumps} and
  \ref{f:seps}.  Below this line there are regions where the nonunit jump $\Delta S^z$ denoted by
  inequality occurs from a state with $S^z > 0$.  The section of the main plot for $\kappa + \beta =
  0.5$ (dotted line) is shown in the inset.  The axis $y$ is the logarithm of the value of the jump
  normed on $\Delta S^z_0 \equiv \Delta S^z |_{\beta = 0}$. \label{f:pt}}
\end{figure}

For magnetization curves of the kind $d$, see Figs.~\ref{f:first-jumps} and \ref{f:jumps}(d), in the
skew phase nonunit jumps are observed, that corresponds to the first-order transition $\Phi_\angle
\leftrightarrow \Phi_\perp$ in the limit case of large spins, but the classical theory predicts that
this transition has a second order, and the region $d$ is absent.  Also the nonequidistantness of
jumps with $\Delta S^z = 1$ in the region of the concentration of jumps (below the bold line in
Fig.~\ref{f:first-jumps}) is observed, see Fig.~\ref{f:jumps}(c).  Their merging is illustrated in
Fig.~\ref{f:jumps}(d).  In the latter case that corresponds to the narrow region $d$ in
Fig.~\ref{f:first-jumps} the magnetization process is the following: several first jumps have
$\Delta S^z = 1$, then the jump with $\Delta S^z > 1$ proceeds (always one), and after that
equidistant jumps with $\Delta S^z = 1$ appear.  In terms of the spin-flop analogue this behavior
can be treated as that the transition $\Phi_\angle \leftrightarrow \Phi_\perp$ is of the first
order, which is impossible for the model given by the expression~\eqref{energy}.  The more detailed
structure of the region $d$ is presented in Fig.~\ref{f:pt} for $2S = 64$.  Parallelograms in the
phase diagram $(b, \beta)$, the region $d$ in Fig.~\ref{f:first-jumps}, corresponds to transitions
from a state with $S^z \neq 0$ to state with $S^z + \Delta S^z$, $\Delta S^z > 1$.  Parallelograms
with $S^z$ and $S^z \pm 1$ is adjoint, and parallelograms with equals $\Delta S^z$ are attached at
corners.  This structure is responsible for the sawtooth shape of curves separating regions with
$\Delta S^z > \rm{const}$ in Fig.~\ref{f:pt}.  However, the isolines in this figure and the plot in
the inset show that the value of the jump decreases rapidly removing from the line of separation
states with $\Delta S^z = 1$ and $\Delta S^z > 1$ in the direction of negative $\beta$ values.

\section{Results and discussion}

Thus, we find out a complicated behavior of the step-like magnetization curves for a pair of quantum
spins with AFM interaction and easy-plane anisotropy that depends on the two anisotropy constants
with different origin.  If the single-ion anisotropy prevails, ''merging'' of unit jumps into the
jump with $\Delta S^z > 1$ can take place.  In the opposite case, when the exchange anisotropy
dominates, all jumps are unit, but they are concentrated in a narrow field range.  We reveal a
qualitative similarity of the quantum problem and the corresponding classical one down to $S \sim
5$.  These observations can be suitable as a starting point for the analysis of the experimental
data ans studying the influence of other kinds of interactions, but it is far beyond this brief
paper.  At the same time, there are quantitative discrepancies between the classical system and the
quantum one.  These discrepancies are important in the range of the fields where the transition from
$\Delta S^z = 1$ to 2 occurs.  The quantum effects that \emph{are free} of a small value $1 / S$
manifest themselves there.  Apparently, the explanation of this unexpected fact is associated with
the follows: for the small anisotropy the quantum model has a singlet ground state that is strongly
dissemble from the quasiclassical Ne\'el picture.  Thus, the region of a pure quantum behavior is
revealed for the system in the limit $S \to \infty$.  Such effects are possible in the physics of
antiferromagnetism.  We mention the Haldane hypothesis\cite{Haldane83} that the ground state of a
quantum spin chain is different for integer and half-integer spin even in the limit $S \to \infty$.
Thus, the simple model of a spin pair considered above demonstrates that antiferromagnets can
manifests substantially quantum properties even for large spins.


\begin{thebibliography}{12}

\bibitem{Wernsdorfer01}
W. Wernsdorfer, Adv. Chem. Phys. {\bf 118}, 99 (2001).

\bibitem{Skomski03}
R. Skomski, J. Phys.: Condens. Matter {\bf 15}, R841 (2003).

\bibitem{Wernsdorfer+02nat}
W. Wernsdorfer, N. Aliaga-Alcalde, D.~N. Hendrickson, and G. Christou, Nature {\bf
 416}, 406 (2002).

\bibitem{LeuenLoss01nat}
M.\,N. Leuenberger and D. Loss, Nature {\bf 416}, 789 (2001).

\bibitem{ShapiraBind02}
Y. Shapira and V. Bindilatti, J. Appl. Phys. {\bf 92}, 4155 (2002).

\bibitem{Wernsdorfer+02prb}
W. Wernsdorfer, S. Bhaduri, C. Boskovic, G. Christou, and D.~N. Hendrickson, 
Phys. Rev. B {\bf 65}, 180403 (2002).

\bibitem{Edwards+03}
R.\,S. Edwards, S. Hill, S. Bhaduri, N. Aliaga-Alcalde, E. Bolina, S. Maccagnano, 
G. Christou, and D. N. Hendrickson, Polyhedron {\bf 22}, 1911 (2003).

\bibitem{Waldmann00}
O. Waldmann, Phys. Rev. B {\bf 61}, 6138 (2000).

\bibitem{BorRom62}
A.\,S. Borovik-Romanov, in {\sl Antiferromagnetism.  Advances in Science} 
(Akad. Nauk SSSR, Moscow, 1962), p.~70 [in Russian].

\bibitem{Belov+79}
K.\,P. Belov, A.\,K. Zvezdin, A.\,M. Kadomtzeva, and R.\,Z. Levitin,
{\sl Orientational transitions in rear-earth magnets} (Nauka, Moscow, 1980) [in Russian].

\bibitem{BarIv85} V.\,G. Bar'yakhtar and B.\,A. Ivanov, in {\sl Intermediate
    state and the dynamic and static properties of domain walls in
    two-sublattice magnets}, Vol.~6 of {\em Sov.  Sci. Rev. Sec. A-Phys.},
  edited by I.\,M.\,Khalatnikov. Harwood, Amsterdam, 1985, pp.\ 404--513.

\bibitem{Haldane83}
F.\,D.\,M. Haldane, Phys. Rev. Lett. {\bf 50}, 1156 (1983).

\end{thebibliography}
\end{document}